\documentclass[aps,superscriptaddress,eqsecnum,nofootinbib,preprintnumbers]{revtex4}
\usepackage{graphicx,epsfig}
\usepackage{amssymb,amsmath}
\usepackage{subfigure}
\usepackage{mwe}
\usepackage{graphicx,epsfig}
\usepackage{amssymb,amsmath}

\usepackage{relsize}

\usepackage[utf8]{inputenc}

\usepackage{xcolor}

\newcommand{\be}{\begin{equation}}
\newcommand{\ee}{\end{equation}}
\newcommand{\ba}{\begin{eqnarray}}
\newcommand{\ea}{\end{eqnarray}}
\newcommand{\beg}{\begin{gather*}}
\newcommand{\eng}{\end{gather*}}

\newcommand{\p}{\partial}

\begin{document}

\title{Constraining  the swiss-cheese IR-fixed point cosmology with cosmic expansion}



\author{Ayan Mitra}
\email{ayan.mitra@nu.edu.kz}
\affiliation{School of Engineering, Nazarbayev University, Republic of Kazakhstan,010000}

\author{Vasilios Zarikas}
\email{vzarikas@uth.gr}
\affiliation{MAE, School of Engineering, Nazarbayev University, Republic of Kazakhstan,010000}

\author{Alfio Bonanno}
\email{alfio.bonanno@inaf.it}
\affiliation{INAF, Osservatorio Astrofisico di Catania,Via S.Sofia 78, IT-95123 Catania, Italy}
\affiliation{INFN, Sezione di Catania,Via S. Sofia 64, IT-95123, Catania, Italy}

\author{Michael Good}
\email{michael.good@nu.edu.kz}
\affiliation{Department of Physics, School of Science and Humanities, Republic of Kazakhstan,010000}

\author{Ertan Güdekli}
\email{gudekli@istanbul.edu.tr}
\affiliation{
Department of Physics, Istanbul University, Istanbul 34134, Turkey }


\begin{abstract}
In a recent work it has been proposed that the recent cosmic passage to a cosmic acceleration era is the result of the existence of small anti-gravity sources in each galaxy and clusters of galaxies.
In particular a swiss-cheese cosmology model which relativistically integrates 
the contribution of all these anti-gravity sources on galactic scale 
has been constructed  assuming the presence of an infrared fixed point for a scale dependent cosmological constant.
The derived cosmological expansion provides explanation for both the fine tuning and the coincidence problem. 
The present work relaxes the previous assumption on the running of the cosmological constant and 
allows for a generic scaling around the infrared fixed point.
Our analysis reveals in order to produce a cosmic evolution consistent with the best $\Lambda$CDM model, 
the IR-running of the cosmological constant is consistent with the presence of an IR-fixed point.
\end{abstract}

\maketitle

\section{Introduction}
The cosmological constant problem of quantum field theory, first emphasized by Zeldovich, is one of the biggest problems of modern theoretical physics. Even if someone would expect quantum gravity to offer an explanation why vacuum does not gravitate (due to non perturbative effects), the natural thing is to have for the present cosmic time, a zero (negligible) cosmological constant; otherwise a fine tuning problem emerges like in the $\Lambda$CDM phenomenological model.

However, the recent cosmic acceleration 
\cite{Riess:1998cb}-\cite{Spergel:2006hy} adds more mystery into the problem.
This missing amount of Dark energy (DE),\cite{Li:2011sd},  perhaps is the result of a time varying cosmological constant $\Lambda(z)$. There are many papers trying to explain this recent low redshifts cosmic acceleration. Among them there are two proposed solutions without the fine tuning problem, work \cite{Zarikas:2011pq} and \cite{Zarikas:2017gfv}.
They remove the coincidence and fine tuning problem connecting the recent large scale structure and its characteristics with the recent cosmic acceleration. The present study concerns a further elaboration of the second work \cite{Zarikas:2017gfv} connecting it and testing it with the cosmological observational data. Note that there are also some other works, \cite{inho},
\cite{inhoswiss}, \cite{structure}, that try to resolve (partially or fully) the recent cosmic acceleration and coincidence problem using the recent large scale structure formation; however these works associate inhomogeneities with the change of the expansion rate. In \cite{Zarikas:2017gfv} local antigravity sources are the reason of the solution and no inhomogeneities play any role. We expect that the inclusion of inhomogeneities will further help the generation of the cosmic acceleration. 

In \cite{Zarikas:2017gfv} the proposed solution is based on
 the infrared quantum gravity modifications at the astrophysical scales. These corrections result to a non zero positive cosmological constant which is related with an astrophysical configuration of matter and energy like galaxies or clusters of galaxies. Thus, effectively antigravity sources are generated. Such quantum gravity corrections perhaps are possible in the framework of the Asymptotic Safety (AS) program
for quantum gravity \cite{ASreviews}, \cite{Falls:2014tra}.

Summing relativistically all homogeneously distributed anti-gravity sources from every galaxy or cluster of galaxies we get a net effect on the cosmic expansion. This "sum" can be done adapting a swiss-cheese modelling in which matching between the local and the cosmic patches generates the observed recent passage from deceleration to acceleration. 

Last ten years many cosmological models have been developed to explain DE. Some of them are more of geometric origin using, i.e. geometric modifications of general relativity, while some of them are considering quantum effects and/or particle physics phenomena and exotic fields. Several are mainly phenomenological solutions while others more close to theoretical concrete models.  Therefore, it is vital to evaluate all of them using observational data, \cite{Jimenez:2001gg}, \cite{Scolnic:2017caz}) . 

The relevant criteria are how well they agree with the characteristics of the late expansion of the universe \cite{newH}. The features of the cosmic history can be revealed using supernovae, SNIa \cite{Suzuki:2011hu}, Gamma Ray Bursts (GRB) \cite{GRB} or HII galaxies \cite{Plionis:2011jj} as luminosity distance estimators.
An alternative way of testing is using the angular diameter distance coming from clusters or CMB sound horizon via Baryon Acoustic Oscillations (BAO) \cite{Blake:2011en}, or utilising the CMB angular power spectrum \cite{Ade:2015xua}. 
Other  probes of the characteristics of the cosmic expansion are 
measures of the growth rate of matter perturbations. 
Galaxies are also very good cosmic
chronometers \cite{Jimenez:2001gg} and the observation of their growth rates  \cite{Basilakos:2017rgc} provide independent of the implied cosmological model knowledge for the the integral of the Hubble parameter $H(z)$.

In \cite{Anagnostopoulos:2018jdq} and \cite{Anagnostopoulos:2019mrc} the cosmological model proposed in  \cite{Zarikas:2017gfv} was tested against observational
data sets: (1) measurements of the Hubble rate $H(z)$, (2) Supernovae
Ia (Pantheon data set, (3) Quasi-Stellar-Objects (QSO), (4) Baryonic Acoustic Oscillations and (5) direct measurements of the CMB shift parameters. The result was that the model was in very good agreement with observations. 
In the present work we instead test the 
possibility of a more general scaling law for the cosmological constant
consistent with the presence of a deviation from the canonical 
scaling which we parametrize by means of an anomalous dimension $\eta$
In particular we  tested the consistency of our model against the backdrop of LCDM cosmology with fiducial values taken from the current observational best fits solutions.

There are two novelties in the present work. In \cite{Zarikas:2017gfv} authors used a phenomenological term for $\Lambda$ while here we point out for the first time that this term can be interpreted as an anomalous dimension due to quantum non-perturbative corrections. Furthermore, in the present work we analyse how sensitive the correct phenomenology of the late cosmology is to the value of $b$. In \cite{Zarikas:2017gfv}, authors claimed that there is no fine tuning in dimensionless parameters. This indeed is true. The present paper proves thoroughly that we don’t have such fine tuning. It is not needed a very precise value of $b$ to get correct behavior.

The structure of the article is as follows: in Sec. \ref{I} we present the theoretical aspects of the Asymptotically Safe swiss-cheese Gravity.
In Sec. \ref{II} we discuss the observational data sets we employ along with our method and the corresponding results. Finally, we summarise our conclusions in Sec. \ref{III}.

\section{swiss-cheese model and IR-fixed point cosmology}
\label{I}

\subsection{Asymptotic Safety in UV and IR}
In the Asymptotic Safety scenario the UV behavior of gravity is ruled by a non-perturbative fixed point
of the renormalization group for quantum gravity \cite{c11}, \cite{c12}. 
The essential ingredient of this approach is the Effective Average Action, a Wilsonian
coarse grained free energy dependent on an infrared momentum scale $k$ which defines an
effective field theory appropriate for the scale $k$. By construction, when evaluated at tree
level, the Effective Average Action correctly describes 
all gravitational phenomena, including all loop effects, if the
typical momentum involved are of the order of $k$. When applied to the Einstein-Hilbert
action the ERG yields renormalization group flow equations which have made possible
detailed investigations of the scaling behavior of Newtons constant at high energies.
The scenario emerging from these studies (see \cite{c2} for a recent discussion on the AS scenario and its
potential application) suggests that the theory could be consistently defined in $d = 4$ 
at a nontrivial UV fixed point where the dimensionless Newton constant, $g(k) = G(k)\, k^2$
does not vanish in the $k\rightarrow \infty$ limit.
As a consequence the dimensionful Newton constant $G(k)$ 
is antiscreened at high energies, very much as one would expect based on the
intuitive picture that the larger is the cloud of virtual particles, the greater is the effective
mass seen by a distant observer \cite{c3}.

Albeit in the original formulation of the AS approach the UV non-gaussian fixed point emerging in the
UV at transplanckian energies to recover the continuum limit, it has been realized that a non-perturbative
IR fixed point at small $k$ could solve the singular behavior of the $\beta$-functions assuming 
that a sort of tree level renormalization could take place at large distances, so that 
\begin{eqnarray}
\label{irfp}
G(k)=g_\ast/k^2 \quad\quad \Lambda(k) = \lambda_\ast k^2 
\end{eqnarray}
for $k\rightarrow 0$.
Clearly $k$ should be considered not as a momentum flowing into a loop, but as an inverse of a typical 
distance over which the averaging of the field variables is performed.  From this point of view
several works have discussed the possibility of interpreting astrophysical data as an infrared 
effect of quantum gravity, relating $k$ to a cosmic time \cite{c4} or cosmic distance. 
In particular in \cite{Zarikas:2017gfv} the running scale $k$ is connected with the typical 
size of a galaxy or of a cluster of galaxies embedded in a swiss-cheese cosmological model,
and for this reason this approach is closer in spirit to the scale identification proposed in 
\cite{Reuter:2004nv,Reuter:2004nx,Reuter:2009kq}. 
In this work we assume that there is no significant running of $G$ at cosmological scale \cite{c5}
but we allow for the possibility of the deviation from the canonical scaling 
in the running of $\Lambda$. In particular we 
set $g_\ast/k^2\sim constant$ and 
$\Lambda(k)=\lambda_\ast k^{2-\eta}$ where $\eta$ is the anomalous dimension, \cite{Bonanno:2011yx}. 
As it is explained in \cite{Bonanno:2011yx}, the varying behaviour of the cosmological constant is based on the instability induced renormalization triggered by the low energy quantum fluctuations in a Universe with a positive cosmological constant.
For $\eta=0$ the canonical (mean-field) scaling is recovered. 

In the following we write $b\equiv 2-\eta$ 
and $b$ has to be determined by assuming an underling best-fit $\Lambda$CDM cosmology.

\subsection{AS swiss-cheese}
This section provides a summary of the mathematical description of the AS inspired swiss-cheese cosmological
model presented in \cite{Zarikas:2017gfv}.
The swiss-cheese model or otherwise called Einstein-Strauss model 
\cite{Einstein:1946ev} describes a global homogeneous and isotropic metric, the Universe, with many
local Schwarzschild black hole metrics homogeneously distributed. The geometrical covariant matching of the global metric as the exterior solution
with a local interior spherical solution happens across a spherical boundary that
is proved to be at a constant coordinate radius of the cosmological metric but at spherical solution's radius evolving in time.

In more detail, we have to geometrically match the exterior background spacetime described by a Friedmann-Lema\^{i}tre-Robertson-Walker (FLRW) metric to an interior local black hole metric. This happens on a spherical 3-surface, $\Sigma$, of constant coordinate radius in the FLRW frame but time evolving in the Schwarzschild frame. 
The matching of the two solutions makes use of the first fundamental form (intrinsic metric) and the second fundamental form
(extrinsic curvature), calculated in terms of the coordinates on $\Sigma$, on both sides,\cite{Israel:1966rt}.

In our case, we use as a local solution, the AS inspired corrected 
Schwarzschild-de Sitter metric. This quantum improved Schwarzschild-de Sitter
metric contains energy dependent cosmological and Newton constants
with the hope to describe fairly astrophysical objects like clusters of galaxies. Thus, in the metric there are energy dependent $G_k=G(k)\sim G_N\,,\,\Lambda_k=\Lambda(k)=\lambda_{*}^{IR}k^{b}$. 
So
\begin{equation}
ds^2=-\Big(1-\frac{2G_k M}{R}-\frac{1}{3}\Lambda_k R^2\Big)dT^2
+\frac{dR^2}{1-\frac{2G_k M}{R}-\frac{1}{3}\Lambda_k R^2}
+R^2\,d\Omega^2\,,
\label{ASBH}
\end{equation}
where now both Newton's constant $G_k$ and cosmological constant $\Lambda_k$ are functions of a
characteristic scale $k$ of the system determined by the quantum gravity theory in use. 

The previous metric should be matched with a homogeneous and
isotropic metric. The cosmological metric is of the form
\begin{equation}
	ds^2=-dt^2+a^2(t)\left[\frac{dr^2}{1-\kappa r^2}+r^2d\Omega^2
	\right]\,,
	\label{eq:FRW}
\end{equation}
where $a(t)$ is the scale factor and $\kappa=0, \pm 1$ characterizes the spatial curvature and $d\Omega^2=\left(d\theta^2+\sin^2\!\theta\,d\varphi^2\right)$ the metric of the two sphere.

We work on a 4-dimensional spacetime $M$ and with a metric $g_{\mu\nu}$. There is a timelike
hypersurface $\Sigma$ that divides $M$ into two regions. The hypersurface $\Sigma$ has induced metric $h_{\mu\nu}=g_{\mu\nu}-n_{\mu}n_{\nu}$, with $n^{\mu}$ the unit normal vector to $\Sigma$ pointing inwards. The extrinsic curvature
is given by $K_{\mu\nu}=h_{\mu}^{\kappa}h_{\nu}^{\lambda}n_{\kappa;\lambda}$. Covariant differentiation with respect to $g_{\mu\nu}$ is denoted by the semicolon "$;$". 
The formalism permits the use of different coordinate systems on both sides of the hypersurface in Darmois-Israel junction conditions \cite{matching}.  Darmois-Israel junction conditions are the two conditions for a smooth matching.
The matching conditions require the continuity of spacetime
across $\Sigma$. The latter means a continuity of the induced metric $h_{ij}$ on $\Sigma$. The Israel-Darmois matching
conditions demand also the sum of the two extrinsic curvatures computed on
the two sides of $\Sigma$ to be zero. 

The first fundamental form is the induced metric metric on $\Sigma$ and is equal to 
\begin{equation}
	\gamma_{\alpha\beta}=g_{ij}\frac{\p x^i}{\p u^\alpha}
	\frac{\p x^j}{\p u^\beta}\, ,
\end{equation}
with $u^\alpha=(u^1\equiv u,\, u^2\equiv v,\, u^3\equiv w)$ 
is the coordinate system
on  $\Sigma$ while $\alpha,\,\,\beta$ take integer values $1,\ldots ,3,$ and english indices $i\,\,,j$ run over
$1,\ldots ,4.$\\
The second fundamental form is given by
\begin{equation}
	K_{\alpha\beta}=n_{i;j}\,\frac{\gamma_{\alpha\beta}}{g_{ij}}=(\Gamma^p{}_{ij}n_p-n_{i,j})
	\frac{\p x^i}{\p u^\alpha}\frac{\p x^j}{\p u^\beta}\, ,
	\label{eq:SFF}
\end{equation}
where $\Gamma^p{}_{ij}$ are the Christoffel symbols.

Note that the cosmic evolution through the swiss-cheese is determined by the cosmic evolution of the matching surface that is happening at sphericial radius, in the black hole frame, $R_{S}$ (and constant $r_\Sigma$ for FLRW frame). This quantity  enters the differential equations. The subscript $S$ is after Schucking and the radius is called Schucking $R_{S}$ radius.

The matching requirements provide the following equations for $R_{S}$,
\begin{eqnarray}
R_{S}&=&ar_{\Sigma}\label{M1}\\
\Big(\frac{dR_{S}}{dt}\Big)^{2}&=&1\!-\!\kappa r_{\Sigma}^{2}-\Big(1-\frac{2G_k M}{R_S}-\frac{1}{3}\Lambda_k R_S^2\Big)
\label{M2}\\
2\frac{d^{2}R_{S}}{dt^{2}}&=&-\frac{d\Big(1-\frac{2G_k M}{R}-\frac{1}{3}\Lambda_k R^2\Big)}{dR}|_{R_{S}}
\label{M3}
\end{eqnarray}

The matching radius $r_{\Sigma}$ can be understood as the boundary of the volume of the interior solid
with energy density equal to the cosmic matter density $\rho$. Thus, the interior energy content should equal the
mass $M$ of the astrophysical object, i.e.
\begin{equation}
r_{\Sigma}=\frac{1}{a_0}\Big(\frac{2G_{\!N}M}{\Omega_{m0}H_{0}^{2}}\Big)^{\!\frac{1}{3}}\,,
\label{lews}
\end{equation}
with density parameter is $\Omega_{m}=\frac{8\pi G_{\!N}\rho}{3H^{2}}=
\frac{2G_{\!N}M}{r_{\Sigma}^{3}a^{3}H^{2}}$ and $a_{0}$ in the following of the study is set to 1.

Eqs. (\ref{M1}, \ref{M2}, \ref{M3}) for constant $G_{k}$ and $\Lambda_{k}$ reduce to the conventional FLRW expansion equations. However, although these equations generate the standard cosmological equations of the
$\Lambda\text{CDM}$ model, even in this case the interpretation is different since the $\Lambda$ term appearing in the black hole metric
(\ref{ASBH}) is like an average of all anti-gravity sources inside the Schucking radius of a galaxy or cluster of galaxies. These antigravity sources one can claim that may arise inside astrophysical black holes in the centers of which the presence of a quantum repulsive pressure could balance the attraction of gravity to avoid the not desired singularity, or it arises due to IR quantum corrections of a concrete quantum gravity theory. Furthermore, for constant $G_{k}$ and $\Lambda_{k}$ the coincidence problem is removed since the large scale structure appears recently.

\subsection{Cosmic acceleration and coincidence problem}
The meaning of the quantity $k$ is understood in a statistical-mechanical sense. The role
of the cutoff $k$ is associated with the block-spin transformation, or Kadanoff blocking, a technique used in lattice field theory and condensed matter \cite{Wilson:1973jj}.
The quantity $k$ is not related with a momentum exchanged in a scattering process of the RG scale often used in dimensional regularization in effective field theory \cite{don}. Assuming Fourier transformability of the block-spin transformation, $k$ is of the order of the inverse of the lattice size.  
Several scaling behaviors for $k$ have been proposed in astrophysical
\cite{Bonanno:2000ep}, \cite{Bonanno:2019ilz} and cosmological contexts
\cite{Bonanno:2007wg}, \cite{2016PhRvD..94j3514K}, \cite{Bonanno:2017pkg}. We use the approach in \cite{Zarikas:2017gfv} where  $k$ has to be associated with a characteristic astrophysical length scale $L$. Thus, 
$k=\xi/L$, with $\xi$ is a dimensionless order-one number. A first choice could be $L$ to be the radial distance $R$ from the center. However this choice proved to be not successful. In \cite{Zarikas:2017gfv} we have shown that for $k=\xi/R$ the behaviour of the coefficient of equation of dark energy state $w_{DE}$ is wrong. There is perhaps a chance to get an acceptable behaviour if one includes redshift dependence of the cluster mass but this remains to be shown.
Another more physically natural option and at the same time generating the correct phenomenology, is to set as $L$ the proper distance $D>0$.  With this choice studies also show a singularity avoidance/smoothing
\cite{Kofinas:2015sna}. 

The proper distance of a radial curve with $dT=d\theta=d\varphi=0$ from $R_0$ till $R$ is estimated by the formula
\begin{equation}
D(R)=\int_{R_{0}}^{R}\frac{d\mathcal{R}}{\sqrt{1-\frac{2G_k M}{\mathcal{R}}-\frac{1}{3}\Lambda_k (\mathcal{R})^2 }}\,.
\label{proper}
\end{equation}

As we have already mentioned the matching happens at the Schucking radius. This means that the relevant value of $k$ is $k_{S}=\xi/D_{S}$, where $D_{S}(R_{S})$
is the proper distance of the Schucking radius.

Using Eqs. (\ref{M1}, \ref{M2}, \ref{M3}) it was shown that, \cite{Zarikas:2017gfv}, the Hubble evolution is given by the system
\begin{eqnarray}
H^{2} & = & -\frac{\kappa}{a^{2}}+\frac{2G_{\!N}M}{r_{\Sigma}^{3}a^{3}}
+\frac{\gamma\xi^{b}}{3D_{S}^{b}} \label{H1} \\
\dot{D}_{S} & = & r_{\Sigma}\,a\,H 
\left(1-\frac{2G_{\!N}M}{r_{\Sigma}a}
-\frac{\gamma\xi^{b}r_{\Sigma}^{2}a^{2}}{3D_{S}^{b}} \right)^{-1/2}
\end{eqnarray}
\label{kewr}
where $\gamma=\lambda_{*}^{IR}$ and the variable $D_{S}$ is of geometrical nature with its own equation of "motion" (time evolution).

One can further prove that the expansion rate of the scale factor is given by 
\begin{equation}
\frac{H^{2}(z)}{H_{0}^{2}}=\Omega_{m0}(1\!+\!z)^{3}
+\Big[\Omega_{DE,0}^{-\frac{1}{b}}-\frac{3^{\frac{1}{b}}}{\xi\tilde{\gamma}^{\frac{1}{b}}}
(G_{\!N}H_{0}^{2})^{\frac{1}{b}}\frac{r_{\Sigma}a_{0}}{\sqrt{G_{\!N}}}\,\frac{z}{1\!+\!z}\Big]^{-b}
+\Omega_{\kappa 0}(1\!+\!z)^{2}\,,
\label{kieb}
\end{equation}
where we have reparameterized $\gamma$ to the dimensionless $\tilde{\gamma}=\gamma\,G_N^{1-b/2}$. 
The acceleration is very well approximated, as before, as a function of $z$ by the expression
\begin{equation}
\frac{\ddot{a}}{H_{0}^{2}a}=-\frac{1}{2}\Omega_{m0}(1\!+\!z)^{3}+\Big[\Omega_{DE,0}^{-\frac{1}{b}}
-\frac{3^{\frac{1}{b}}}{2\xi\tilde{\gamma}^{\frac{1}{b}}}(G_{\!N}H_{0}^{2})^{\frac{1}{b}}
\frac{r_{\Sigma}a_{0}}{\sqrt{G_{\!N}}}\frac{b\!+\!2z}{1\!+\!z}\Big]
\Big[\Omega_{DE,0}^{-\frac{1}{b}}
-\frac{3^{\frac{1}{b}}}{\xi\tilde{\gamma}^{\frac{1}{b}}}(G_{\!N}H_{0}^{2})^{\frac{1}{b}}
\frac{r_{\Sigma}a_{0}}{\sqrt{G_{\!N}}}\frac{z}{1\!+\!z}\Big]^{-1-b}\,.
\label{ther}
\end{equation}
and the equation of state parameter is 
\begin{equation}
w_{DE}=\Big[\frac{3^{\frac{1}{b}-1}}{\xi\tilde{\gamma}^{\frac{1}{b}}}(G_{\!N}H_{0}^{2})^{\frac{1}{b}}
\frac{r_{\Sigma}a_{0}}{\sqrt{G_{\!N}}}\frac{b\!+\!3z}{1\!+\!z}-\Omega_{DE,0}^{-\frac{1}{b}}\Big]
\Big[\Omega_{DE,0}^{-\frac{1}{b}}
-\frac{3^{\frac{1}{b}}}{\xi\tilde{\gamma}^{\frac{1}{b}}}(G_{\!N}H_{0}^{2})^{\frac{1}{b}}
\frac{r_{\Sigma}a_{0}}{\sqrt{G_{\!N}}}\frac{z}{1\!+\!z}\Big]^{-1}\,
\label{wde}
\end{equation}

The present study will test Equations (\ref{kieb}, \ref{wde})  against observations. The free parameters are $b$, $\xi,\, \tilde{\gamma}$. The values of  $\xi$, and $\tilde{\gamma}$ are not known but it is expected to be of order O(1).
The crucial one is $b$ since it is related with the anomalous dimension that arises due to non perturbative quantum corrections; and so the analysis will focus on this. The model does not have many free parameters and it is remarkable that provides the correct phenomenology. The way we have chosen the values of the free parameters is as follows: One can choose any value of order one for $\xi$ and $\tilde{\gamma}$ and then $b$ must have a value that ensures that the second term and the third term of Eq.(\ref{H1}) are almost equal. There must be a balance between attraction and repulsion for low redshifts. Selecting other values for  $\xi$ and $\tilde{\gamma}$ changes slughtly the value of $b$ for correct phenomenology.
Note also, that not only the proper distance today $D_{S,0}$, but also the whole function $D_{S}(z)$ remains for all observational tests range of redshift $z$, of the order of $r_{\Sigma}$.

\section{Methodology and Numerical results}
\label{II}

For comparison of our model, with $\Lambda$CDM cosmology, we consider a flat universe, i.e. $\Omega_k=0$. Therefore, the Hubble parameter in eq.\ref{kieb} can be re written as,
\begin{equation}
{H^{2}(z)}={H_{0}^{2}\left[\Omega_{m0}(1\!+\!z)^{3}
+\Big[\Omega_{DE,0}^{-\frac{1}{b}}-\frac{3^{\frac{1}{b}}}{\xi\tilde{\gamma}^{\frac{1}{b}}}
(G_{\!N}H_{0}^{2})^{\frac{1}{b}}\frac{r_{\Sigma}a_{0}}{\sqrt{G_{\!N}}}\,\frac{z}{1\!+\!z}\Big]^{-b}\right]}\,
\label{kieb2}
\end{equation}
In our analysis for comparing and estimating the optimal range for the free parameter of our model, we assumed the best fit values of the involved parameters as fiducial values, this includes $\Theta=\{\Omega_m^0,\,H_0^{},\,w_{DE}^{}\}$  computed at their fiducial values $\Theta_\mathrm{fid}=\{0.315,\,67.4,\,-1\}$. The current observational best fit value of $H_0$ is $67.4$ $Km\,s^{-1}Mpc^{-1}$ adapted from the latest CMB inferred constraints from the Planck collaboration \cite{planck} while dark energy equation of state parameters are motivated from the baseline fiducial values of SNe Ia observations.
We have chosen Planck collaboration data, because it has been one of the most popular cosmological probes, providing us the tightest constraints. However, note that there are also other data coming from local observations that suggest somewhat different values \cite{local}.

\begin{figure}
\centering
\begin{minipage}{.45\linewidth}
  \includegraphics[width=\linewidth]{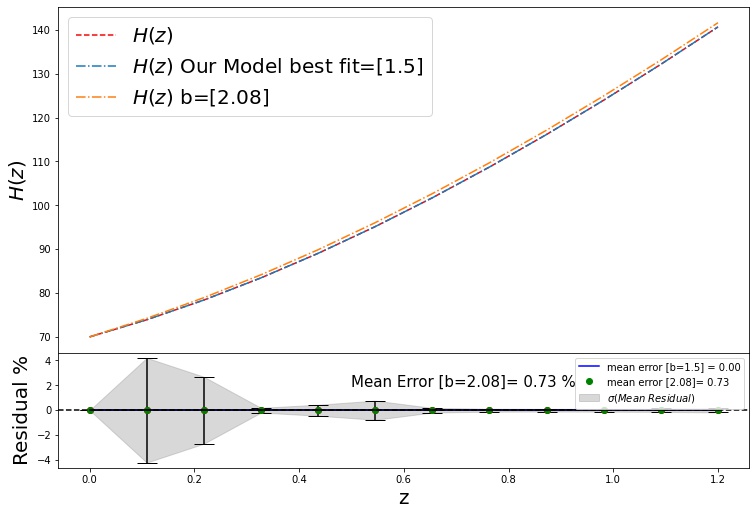}
  \label{img111}
\end{minipage}
\hspace{.05\linewidth}
\begin{minipage}{.45\linewidth}
  \includegraphics[width=\linewidth]{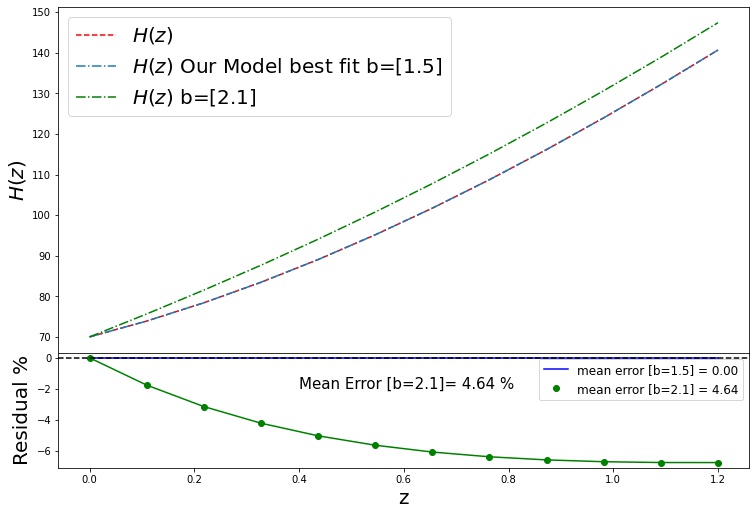}
  \label{img222}
\end{minipage}
\caption{\textbf{Left} : Comparison of $H(z)$ between $\Lambda$CDM values  and our model's values with best fit  value of $b=1.505$ and for $b=2.08$ which corresponds to the maximum value of $b$ for the residual error to be $\le1\%$. The error bars in the residue plot panel (and the grey highlighted zone), corresponds to the standard deviation in the respective redshift bin, computed from all the $b$ values mentioned in the main text. \textbf{Right} : The same comparison, but for $b=2.11$, which corresponds to the error limit of $\le 5\%$. The bottom panel in both the plots show the corresponding residual plot. }
\label{f1111}
\end{figure}

Our expression of the Hubble parameter is a function of the $b$ parameter apart from the redshift $(z)$. We analysed our model to evaluate the allowed range of $b$, to match the observational results. While using Eq.(\ref{kieb2}), we first have designed our swiss-cheese model to model a typical cluster of galaxies with a De-Sitter Schwarzchild black hole. After we will also commend about the matching at the galaxy scale. Based on this assumptions, we chose the values of $\{\tilde{\gamma},\xi\}\ =\ [5.0,9.0]$. Further, the matching radius ($r_{\Sigma}$) is chosen to be $18 Mpc$ for the current epoch where $a_0=1$. For this, we considered a redshift range of $z=[0-1.2]$ and computed the Hubble parameter ($H(z,b)$) for a range of $b$ for the entire redshift range. We chose the range of $b=[0,3.0]$.  We then computed the mean error for each $b$ with the $\Lambda$CDM Hubble parameter.  Based on the mean error  values of $b$, we set two different cut off levels, $[1\%,\ 5\%]$ to estimate the range of $b$ values, we would rate as permissible for application. Additionally we performed a parameter fitting analysis, to estimate the best fit $b$ value for the Hubble parameter. The best fit value obtained is $b=1.505\ \sim1.5$. We found that for keeping the overall residual deviation error between the two Hubble parameter values to under $1\%$ level, the maximum permissible range of $b$ is $2.08$, whereas if the error level considered is up to $5 \%$ margin, then the maximum $b=2.11$. This is illustrated in the two plots in Figure \ref{f1111}.

\begin{figure}
\centering
\begin{minipage}{.45\linewidth}
  \includegraphics[width=\linewidth]{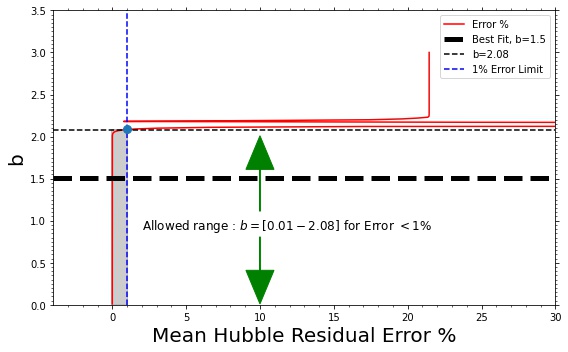}
  \label{img1}
\end{minipage}
\hspace{.05\linewidth}
\begin{minipage}{.45\linewidth}
  \includegraphics[width=\linewidth]{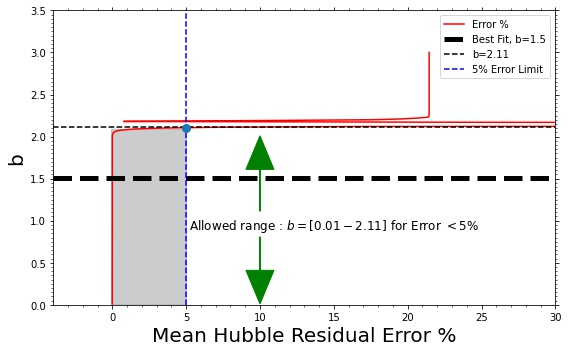}
  \label{img2}
\end{minipage}
\caption{\textbf{Left} : 1\% error. \textbf{Right} : 5\% error limit. The red line shows the error percentage against the $b$ value. The blue dashed horizontal line marks the two corresponding level of error margin.  The grey shaded region marks the allowed range of $b$ values for which the error level will be either $\le1\%$ or $\le5\%$. The horizontal black dashed line, marks the maximum limit of the $b$ value. The thick black horizontal line shows the best fit $b$ value.}
\label{f1222}
\end{figure}

Figure \ref{f1222} shows the allowed $b$ range within the specified error levels of $1\%$ and $5\%$. The margin of $b$ obtained from these results are summarised in the table below,

\begin{figure}
\centering
\begin{minipage}{.45\linewidth}
  \includegraphics[width=\linewidth]{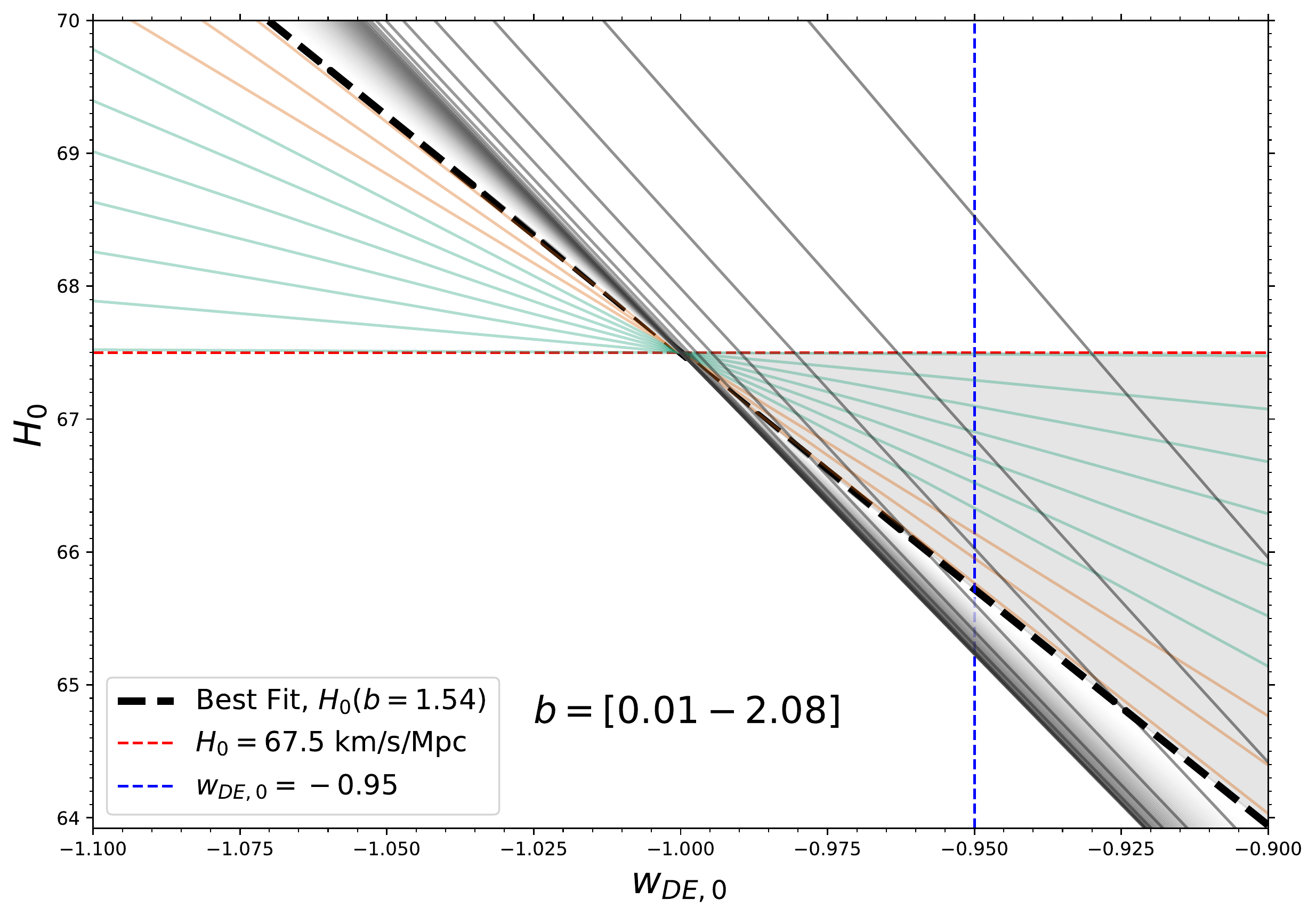}
  \label{img11}
\end{minipage}
\hspace{.05\linewidth}
\begin{minipage}{.45\linewidth}
  \includegraphics[width=\linewidth]{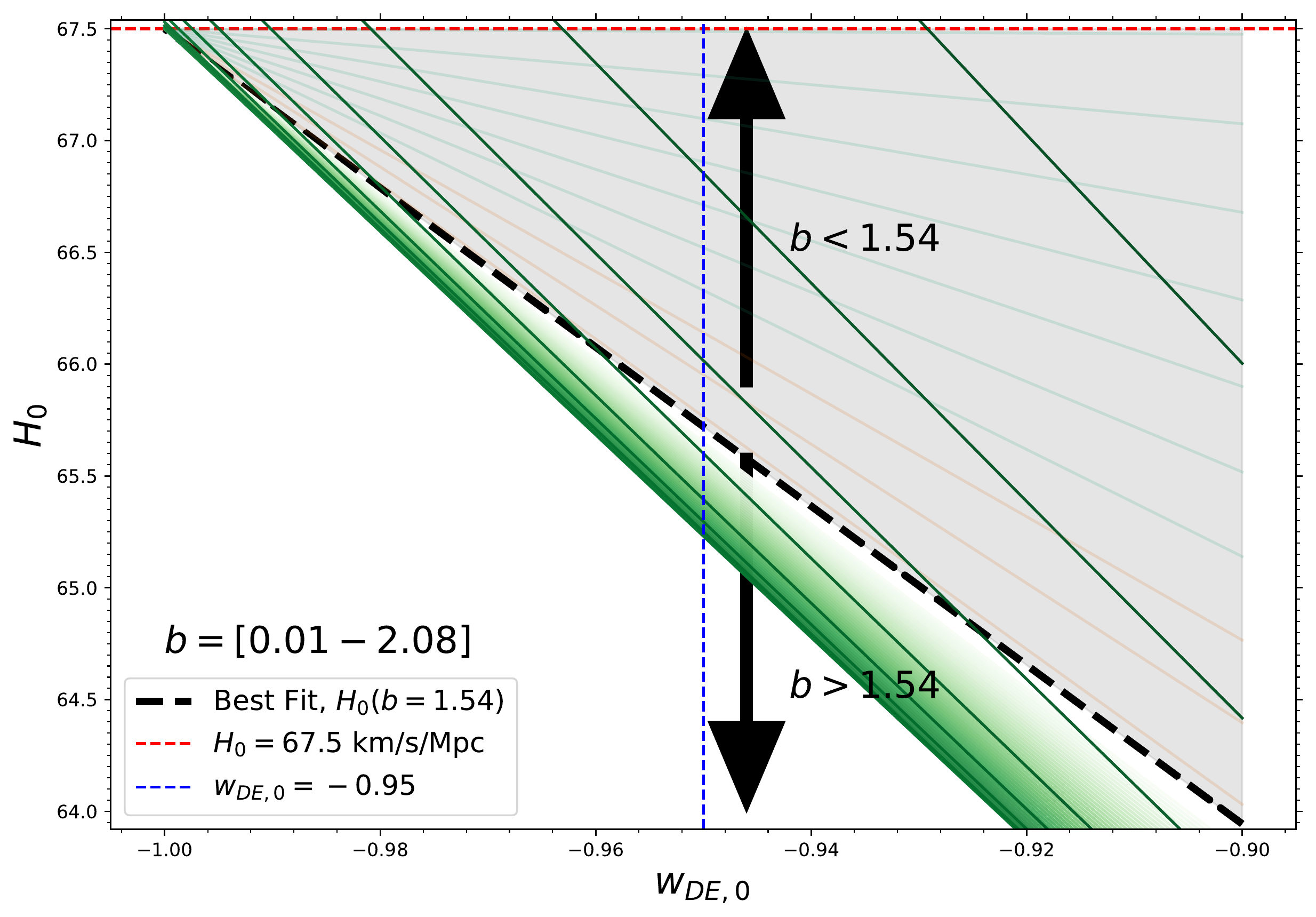}
  \label{img22}
\end{minipage}
\caption{\textbf{Left} : Value of $H_0$ plotted against the current day $w_{DE}$ values for a family of $b$ values. The lines are demarcated into two broad groups, one $b<1.54$ (coloured lines) the other $b>1.54$ (grey tone lines), seperated by the $b=1.54$ line, shown in broad black dashed line, which corresponds to the best fit $b$ value.  \textbf{Right} : A zoom in plot of the left figure. the corresponding grey toned lines (for $b>1.54$ case) are shown in green tone lines here. It is observed that for low $b$ values the evolution of Hubble constant value at present time has minimum correlation to the current dark energy parameter constraint. As $b$ is increased the tilt in the $H_0-w_{DE,0}$ relation increases towards a more positive correlation. The blue vertical dashed line marks the $w_{DE,0}$ value from our model. As we see, that for very low values of $b$ or for $b>2.065$ the $H_0$ values come closest to the current observed $H_0=67.5$ $km/s/Mpc$ value.}
\label{f1223}
\end{figure}

It is interesting to note that Eq.(\ref{kieb2}) (i.e. for a flat Universe), can be related to the Eq.(\ref{wde}), the combined expression is as follows,
\begin{equation}
H^2(z) = H_0^2\left[\Omega_m (1+z)^3 + \left(\frac{w_{DE}}{\left[\frac{3^{(\frac{1}{b}-1)}}{\xi\tilde{\gamma}^{\frac{1}{b}}}\left(G_{\!N}H_{0}^{2}\right)^{\frac{1}{b}}
\frac{r_{\Sigma}a_{0}}{\sqrt{G_{\!N}}}\frac{b+3z}{1\!+\!z} -\Omega_{DE,0}^{-\frac{1}{b}}   \right]}\right)^b\right]
\label{h_w}
\end{equation}

As a result, this allowed us to further check the consistency of the Hubble parameter values, as a function of $b$, with the current value of the equation of state parameter $w_{DE,0}=-0.95$ obtained from our model, as mentioned before.

Based on the value of $H(z)$, we made a best fit analysis with the $H(z)$ from $\Lambda$CDM expression, and obtained the $b$ best fit. It is seen that from comparing with the Hubble parameter solely, the best fit obtained is $b=1.54$.  An intuitive check is performed by comparing the $H_0$ values with the $w_{DE,0}$ values for a family of  $b$ values shown in Fig.\ref{f1222}.  it is seen that the for $b>2.065$ the observed value of $H_0$ is closest to the current observed data of $H_0=67.5$ $km\,s^{-1}\,Mpc^{-1}$.

\subsection{$b$ Constraint From $w_{DE,0}$ and $H_0$ Combination}
Additionally we checked the residual error on $H_0$ with current observational value  while asserting the current dark energy equation of state $w_{DE,0}$ value from our model, i.e. $w_{DE,0}=-0.95$ in Eq.(\ref{h_w}). From this analysis, the minimum residual error  obtained corresponds to $b=2.065$, this is illustrated in the Figure 4,
\begin{figure}[ht]
\centering
\includegraphics[width=10cm]{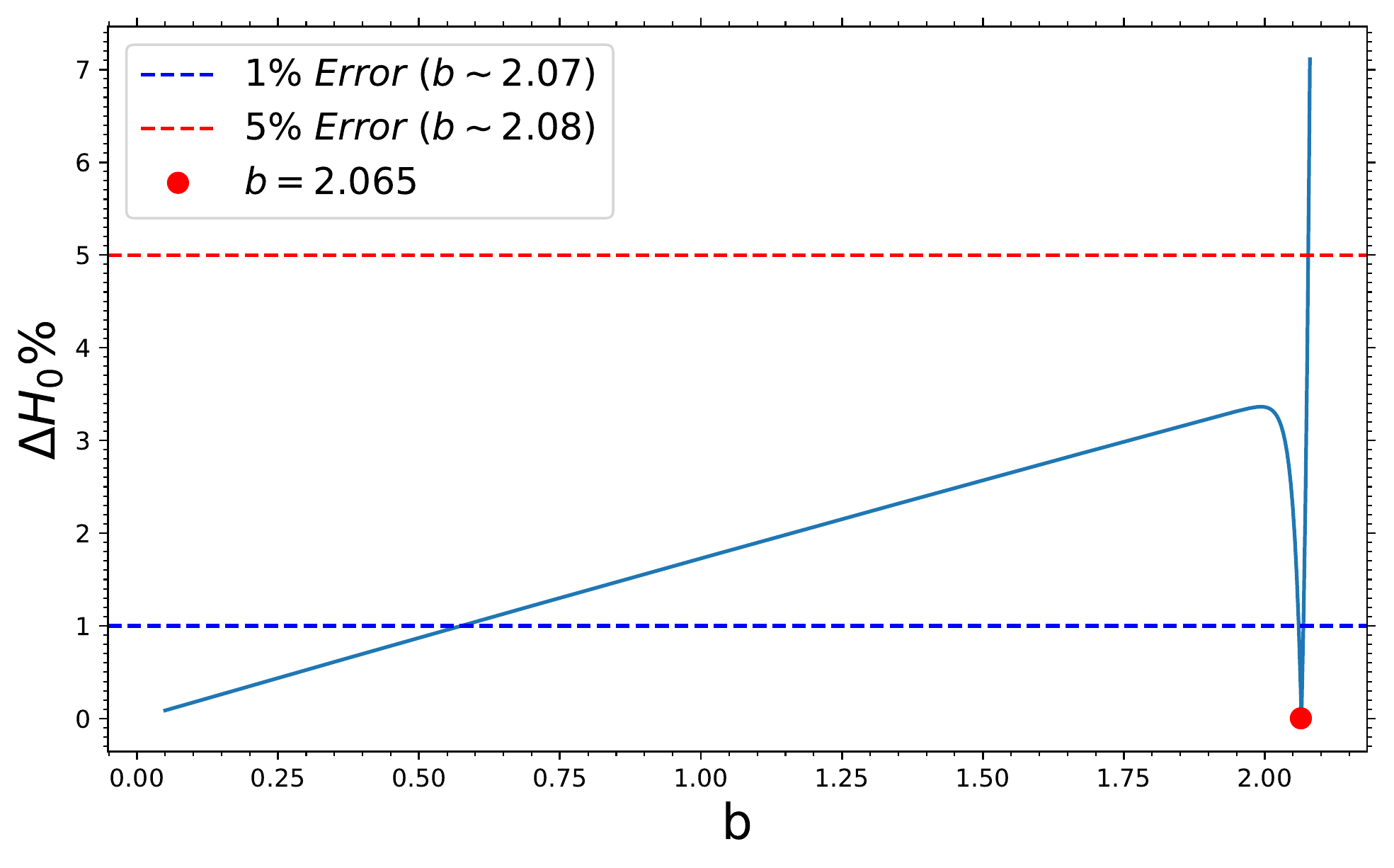}
\caption{Residual plot of $\Delta H_0$ computed from the difference of 67.5 km/s/Mpc and $H_0$ computed at $z=0,w_{DE,0}=-0.95$ from the  eq.\ref{h_w} for a family of `$b$' values. As seen, the red dot corresponding to $b\simeq2.07$  marks the point of lowest residual. The blue and the red dashed horizontal line marks the $1\%$ and $5\%$ error levels.}
\end{figure}
while the $1\%$ and $5\%$ error levels are marked with horizontal lines and their corresponding $b$ values. 
\begin{figure}[ht]
\centering
\includegraphics[width=10cm]{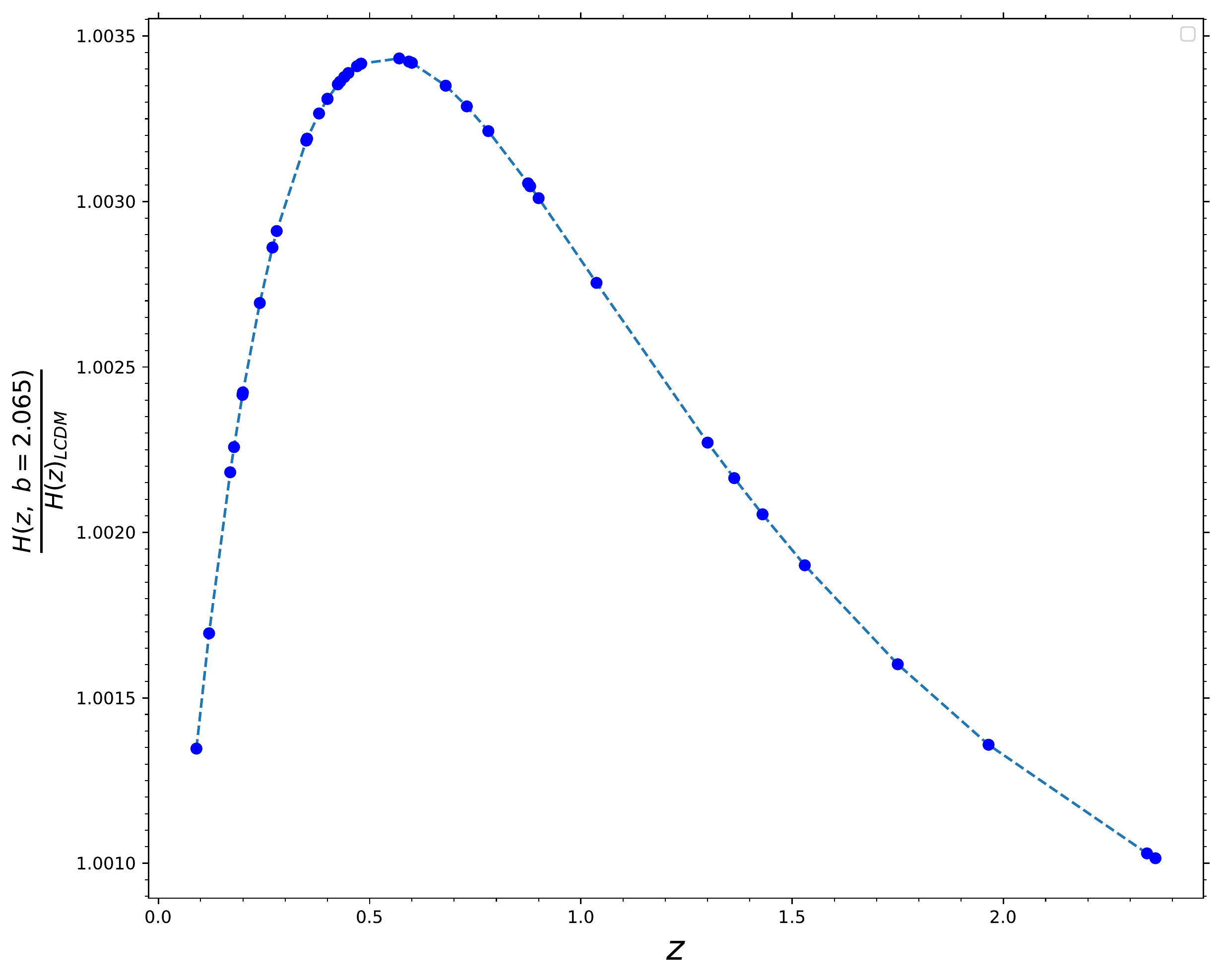}
\caption{Plot showing the ratio between the Hubble parameter ($H(z)$) from our Swiss cheese model computed at the best fit value of $b=2.065$ and the Hubble parameter corresponding to the LCDM model, for the redshift ranges of $z=[0-2.4$]. } \label{f55}
\end{figure}
Thus we find that applying the joint constraints of dark energy $w_{DE}$ and the Hubble constant $H(z)$ we can arrive at a tighter requirement  on the acceptable range of values for $b$. Additionally in  fig.\ref{f55} we see that the best fit Swiss cheese model $H(z)$ are in excellent agreement with the $\Lambda$CDM model.

  \begin{table}
        \begin{center}
            \begin{tabular}{ c c c c }
                \hline
                Error \% & $b$ (from $H(z)$) &  $b$ (from $H0+w_{DE,0}$)\\[.3em]
                \hline\hline
                1 & 2.08 & 2.07\\ 
                5 & 2.11 & 2.08\\ 
                \hline
            \end{tabular}
        \end{center}
        \caption{The upper limit of $b$ for the corresponding permissible error margin, computed from the Hubble parameter estimates and Hubble constant and $w_{DE,0}$ combined constraint. }
        \label{T1}
    \end{table}

\subsection{Observational Hubble Data}
\begin{figure}[ht]
\centering
\includegraphics[width=16cm]{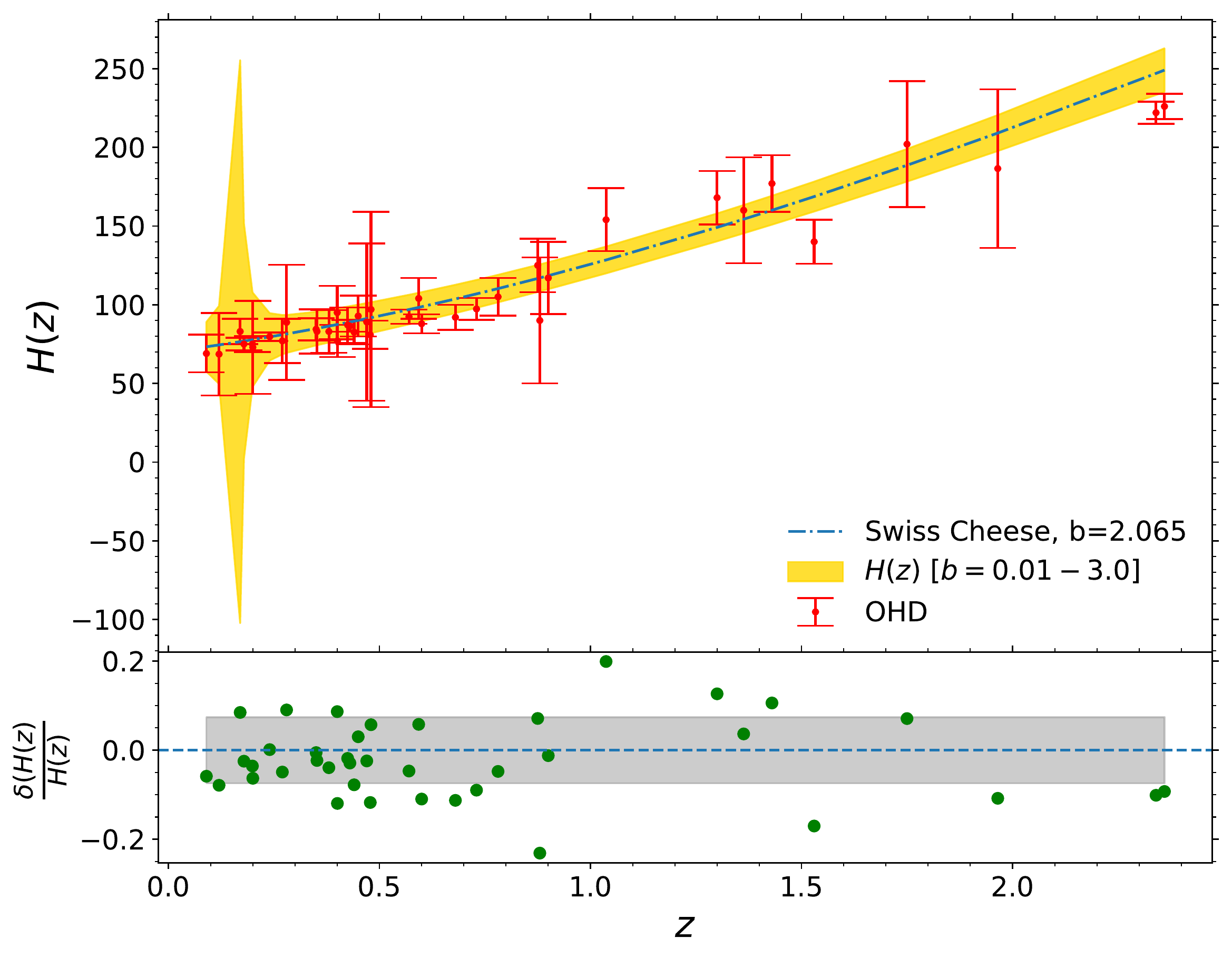}
\caption{Plot of the best fit Hubble parameter value of our Swiss cheese model, $H(z,\ b=2.065)$} (green dashed) against the redshift binned OHD with their corresponding error bars (red) for $z=[0-2.4]$. The yellow margin represents the extent of the $H(z,b)$ values of our model when $b$ is varied between $0.01-3.0$ across the full redshift range. The bottom plot shows the ratio of the error between the OHD data (red in top panel) and the best fit Swiss cheese data (green in top panel) i.e. $\delta (H)$ and the Hubble parameter (green dots). The grey shaded margin corresponds to the mean error level $(\sim 0.07)$.
\label{f66}
\end{figure}

Finally we match our theory with observational Hubble data (OHD) obtained from two additional cosmic probe methods, the baryon acoustic oscillation (BAO) and the cosmic chronometer, coupled across the redshift range of $z=[0-2.4]$ in $40$ redshift bins. We derived the OHD data from \cite{OHD}, who constructed it from various BAO and cosmic chronometer observational data mentioned therein \cite{OHD1,OHD2,OHD3,OHD4,OHD5,OHD6,OHD7,OHD8,OHD9,OHD10,OHD11,OHD12,OHD13,OHD14}. As it can be seen our model at its best fit $'b'$ value gives a very agreement to the observational data collected from the BAO and the cosmic chronometers together as shown in fig.\ref{f66}, across almost the entire redshift range. 

\subsection{Matching at the galaxy length scale }
Following the same method, we have also studied Swiss-cheese model with matching at the galaxy scale. We found that the constraints are more relaxed at the galaxy level. We used the corresponding parameter values $\{ \tilde\gamma, \xi\} \ =\ [5.0,9.0]$ which is same as the cluster swiss-cheese model configuration but with a different matching radius, $r_{\Sigma}=0.83Mpc$. Carrying out an exact similar analysis (like shown in fig.\ref{f1223}), showed that the permissible range for $b$ is  $(0.01-2.13)$ for deviation with observational constraints to be $<1\%$ level, while for the $5\%$ level the upper limit of the constraint goes up to $2.16$ when considering only Hubble constant constraints. When combined with the dark energy ($w_{DE}$) constraint, the corresponding margins for $b$ are tightened slightly, as listed in table \ref{T2}.
 \begin{table}
        \begin{center}
            \begin{tabular}{ c c c c }
                \hline
                Error \% & $b$ (from $H(z)$) &  $b$ (from $H0+w_{DE,0}$)\\[.3em]
                \hline\hline
                1 & 2.13 & 2.11\\ 
                5 & 2.16 & 2.13\\ 
                \hline
            \end{tabular}
        \end{center}
        \caption{Same as in table \ref{T1} but with the Galaxy based swiss-cheese model. }
        \label{T2}
    \end{table}

However, this second case of matching at the galactic scale is not a favorable one. The reason is that swiss-cheese modeling at the galaxy scale is less realistic description since the astrophysical typical galactic length is not so close to the matching Shucking radius like in the clusters of galaxies case.


\section{Conclusions}
\label{III}

In this work we study the phenomenology of a recently proposed AS inspired, IR-fixed point swiss-cheese model, \cite{Zarikas:2017gfv}. In this model a spatially homogeneous isotropic universe, is filled with a big number of quantum improved Schwarzschild de-Sitter black holes uniformly distributed throughout the space. Each such sphere can be physically realized by an astrophysical object, such as a galaxy or a cluster of galaxies (with its extended spherical halo).

The tests with observations consolidate the belief that the model is extremely well behaved in terms of current observational fits, when the free parameter `b' is set within the proper range. We showed using results from a combination of different cosmic probes data, that with the best fit $'b'$ value we can fit the observational data, especially the Hubble parameter to a very good agreement, one which can mimick the LCDM behaviour to very accuracy as shown in the comparison plot in fig.\ref{f55}. Fine tuning is not required. 
In future it will be interesting to apply  
further constraints from upcoming future survey data \cite{sn1, sn2} and other probes like clustering, FRB \cite{FRB}, gravitational waves \cite{gw} etc. and examine the allowed range for the anomalous dimension $eta$ which is related to the free parameter parameter $b=2-\eta$ in model \cite{Zarikas:2017gfv}. 

We have successfully presented a strong case for an alternate model to the traditional $\Lambda$CDM cosmology which is consistent with our current observations and is well suited to explain the recent cosmic acceleration and its coincidence problem in a minimal way without exotic field and without fine tuning or extra scales. The simplicity of this model and its feasibility makes it a strong candidate for upcoming future surveys to further follow up on it's framework.  

As a future work it would be interesting to explore the phenomenological properties of the AS inspired swiss-cheese IR point cosmological model along the lines of works \cite{future}.

\section{Acknowledgments}
The authors A.M, V.Z and A.B acknowledge the support of the Faculty Development Competitive Research Grant Program (FDCRGP) Grant No. 110119FD4534.


\begin{thebibliography}{0}
\expandafter\ifx\csname natexlab\endcsname\relax\def\natexlab#1{#1}\fi
\expandafter\ifx\csname bibnamefont\endcsname\relax
  \def\bibnamefont#1{#1}\fi
\expandafter\ifx\csname bibfnamefont\endcsname\relax
  \def\bibfnamefont#1{#1}\fi
\expandafter\ifx\csname citenamefont\endcsname\relax
  \def\citenamefont#1{#1}\fi
\expandafter\ifx\csname url\endcsname\relax
  \def\url#1{\texttt{#1}}\fi
\expandafter\ifx\csname urlprefix\endcsname\relax\def\urlprefix{URL }\fi
\providecommand{\bibinfo}[2]{#2}
\providecommand{\eprint}[2][]{\url{#2}}

\end{thebibliography}


\begin{thebibliography}{99}
	
	
	
	\bibitem{Riess:1998cb}
	A.~G.~Riess {\it et al.} [Supernova Search Team],
	Astron.\ J.\  {\bf 116}, 1009 (1998)
	doi:10.1086/300499
	[astro-ph/9805201].
	
	
	\bibitem{Perlmutter:1998np}
	S.~Perlmutter {\it et al.} [Supernova Cosmology Project Collaboration],
	Astrophys.\ J.\  {\bf 517}, 565 (1999)
	doi:10.1086/307221
	[astro-ph/9812133].
	
	
	\bibitem{Knop:2003iy}
	R.~A.~Knop {\it et al.} [Supernova Cosmology Project Collaboration],
	Astrophys.\ J.\  {\bf 598}, 102 (2003)
	doi:10.1086/378560
	[astro-ph/0309368];
	A.~G.~Riess {\it et al.} [Supernova Search Team],
	Astrophys.\ J.\  {\bf 607}, 665 (2004)
	doi:10.1086/383612
	[astro-ph/0402512].
	
	
	\bibitem{Spergel:2006hy}
	D.~N.~Spergel {\it et al.} [WMAP Collaboration],
	Astrophys.\ J.\ Suppl.\  {\bf 170}, 377 (2007)
	doi:10.1086/513700
	[astro-ph/0603449].
	
	
	\bibitem{Li:2011sd}
	M.~Li, X.~D.~Li, S.~Wang and Y.~Wang,
	Commun.\ Theor.\ Phys.\  {\bf 56}, 525 (2011)
	doi:10.1088/0253-6102/56/3/24
	[arXiv:1103.5870 [astro-ph.CO]].
	
	\bibitem{Zarikas:2011pq} 
	G.~Kofinas and V.~Zarikas,
	Eur.\ Phys.\ J.\ C {\bf 73}, no. 4, 2379 (2013)
	doi:10.1140/epjc/s10052-013-2379-9
	[arXiv:1107.2602 [hep-th]].
	
	
	\bibitem{Zarikas:2017gfv}
	G.~Kofinas and V.~Zarikas,
	Phys. Rev. D \textbf{97} (2018) no.12, 123542
	doi:10.1103/PhysRevD.97.123542
	[arXiv:1706.08779 [gr-qc]].
	
	
	\bibitem{inho}
	G.~F.~R.~Ellis and W.~Stoeger,
	Class.\ Quant.\ Grav.\  {\bf 4}, 1697 (1987)
	doi:10.1088/0264-9381/4/6/025; 
	S.~R.~Green and R.~M.~Wald,
	Phys.\ Rev.\ D {\bf 83}, 084020 (2011)
	doi:10.1103/PhysRevD.83.084020
	[arXiv:1011.4920 [gr-qc]]; 
	T.~Buchert, M.~J.~France and F.~Steiner,
	Class.\ Quant.\ Grav.\  {\bf 34}, no. 9, 094002 (2017),
	doi:10.1088/1361-6382/aa5ce2
	[arXiv:1701.03347 [astro-ph.CO]]; 
	T.~Buchert, A.~A.~Coley, H.~Kleinert, B.~F.~Roukema and D.~L.~Wiltshire,
	Int.\ J.\ Mod.\ Phys.\ D {\bf 25}, no. 03, 1630007 (2016),
	doi:10.1142/S021827181630007X, 10.1142/9789813226609-0034
	[arXiv:1512.03313 [astro-ph.CO]].
	
	\bibitem{inhoswiss}
	S.~M.~Koksbang,
	Phys.\ Rev.\ D {\bf 95}, no. 6, 063532 (2017)
	doi:10.1103/PhysRevD.95.063532
	[arXiv:1703.03572 [astro-ph.CO]];
	K.~Bolejko and M.~N.~Celerier,
	Phys.\ Rev.\ D {\bf 82}, 103510 (2010)
	doi:10.1103/PhysRevD.82.103510
	[arXiv:1005.2584 [astro-ph.CO]];
	P.~Mishra, M.~N.~Celerier and T.~P.~Singh,
	Phys.\ Rev.\ D {\bf 86}, 083520 (2012)
	doi:10.1103/PhysRevD.86.083520
	[arXiv:1206.6026 [astro-ph.CO]];
	T.~Biswas and A.~Notari,
	JCAP {\bf 0806}, 021 (2008)
	doi:10.1088/1475-7516/2008/06/021
	[astro-ph/0702555];
	V.~Marra, E.~W.~Kolb and S.~Matarrese,
	Phys.\ Rev.\ D {\bf 77}, 023003 (2008)
	doi:10.1103/PhysRevD.77.023003
	[arXiv:0710.5505 [astro-ph]].
	
	\bibitem{structure}
	S.~Rasanen,
	EAS Publ.\ Ser.\  {\bf 36}, 63 (2009)
	doi:10.1051/eas/0936008
	[arXiv:0811.2364 [astro-ph]];
	S.~Rasanen,
	arXiv:1012.0784 [astro-ph.CO];
	R.~A.~Sussman,
	Class.\ Quant.\ Grav.\  {\bf 28}, 235002 (2011)
	doi:10.1088/0264-9381/28/23/235002
	[arXiv:1102.2663 [gr-qc]];
	M.~Lavinto, S.~Rasanen and S.~J.~Szybka,
	JCAP {\bf 1312}, 051 (2013)
	doi:10.1088/1475-7516/2013/12/051
	[arXiv:1308.6731 [astro-ph.CO]].
	
	
	
	
	\bibitem{ASreviews}
	M.~Niedermaier and M.~Reuter,
	Living Rev.\ Rel.\  {\bf 9}, 5 (2006);
	R.~Percacci,
	In {\textit{Oriti, D. (ed.): Approaches to quantum gravity}} 111-128
	[arXiv:0709.3851 [hep-th]];
	O.~Lauscher and M.~Reuter,
	In {\textit{Fauser, B. (ed.) et al.: Quantum gravity}} 293-313 [hep-th/0511260];
	M.~Reuter and F.~Saueressig,
	New J.\ Phys.\  {\bf 14}, 055022 (2012)
	[arXiv:1202.2274 [hep-th]];
	A.~Bonanno,
	PoS CLAQG {\bf 08}, 008 (2011) [arXiv:0911.2727 [hep-th]];
	M.~Niedermaier,
	Class.\ Quant.\ Grav.\  {\bf 24}, R171 (2007) [gr-qc/0610018];
	R. Percacci, {\textit{``An Introduction to Covariant Quantum Gravity and Asymptotic Safety''}},
	Word Scientific, ISBN: 978-981-3207-17-2;
	R. Percacci and D. Perini, Phys. Rev. D68, 044018 (2003).
	
	
	\bibitem{Falls:2014tra}
	K.~Falls, D.~F.~Litim, K.~Nikolakopoulos and C.~Rahmede,
	Phys.\ Rev.\ D {\bf 93}, no. 10, 104022 (2016)
	doi:10.1103/PhysRevD.93.104022
	[arXiv:1410.4815 [hep-th]];
	
	
	\bibitem{Jimenez:2001gg}
	R.~Jimenez and A.~Loeb,
	Astrophys.\ J.\  {\bf 573}, 37 (2002)
	doi:10.1086/340549
	[astro-ph/0106145].
	
	\bibitem{Scolnic:2017caz}
	D.~M.~Scolnic {\it et al.},
	Astrophys.\ J.\  {\bf 859} (2018) no.2,  101
	doi:10.3847/1538-4357/aab9bb
	[arXiv:1710.00845 [astro-ph.CO]];
	The numerical data of the full Pantheon SnIa sample are available at
	http://dx.doi.org/10.17909/T95Q4X
	https://archive.stsci.edu/prepds/ps1cosmo/index.html.
	
	\bibitem{newH}
A.~L.~Ratsimbazafy, S.~I.~Loubser, S.~M.~Crawford, C.~M.~Cress, B.~A.~Bassett, R.~C.~Nichol and P.~V\"ais\"anen,
Mon. Not. Roy. Astron. Soc. \textbf{467} (2017) no.3, 3239-3254
doi:10.1093/mnras/stx301
[arXiv:1702.00418 [astro-ph.CO]].
;
M.~Moresco, L.~Pozzetti, A.~Cimatti, R.~Jimenez, C.~Maraston, L.~Verde, D.~Thomas, A.~Citro, R.~Tojeiro and D.~Wilkinson,
JCAP \textbf{05} (2016), 014
doi:10.1088/1475-7516/2016/05/014
[arXiv:1601.01701 [astro-ph.CO]].
	
	\bibitem{Suzuki:2011hu}
	N.~Suzuki {\it et al.},
	Astrophys.\ J.\  {\bf 746}, 85 (2012)
	doi:10.1088/0004-637X/746/1/85
	[arXiv:1105.3470 [astro-ph.CO]];
	
	
	
	\bibitem{GRB}
	L.~Amati {\it et al.},
	Astron.\ Astrophys.\  {\bf 390}, 81 (2002)
	doi:10.1051/0004-6361:20020722
	[astro-ph/0205230];
	G.~Ghirlanda, G.~Ghisellini and C.~Firmani,
	New J.\ Phys.\  {\bf 8}, 123 (2006)
	doi:10.1088/1367-2630/8/7/123
	[astro-ph/0610248];
	S.~Basilakos and L.~Perivolaropoulos,
	Mon.\ Not.\ Roy.\ Astron.\ Soc.\  {\bf 391}, 411 (2008)
	doi:10.1111/j.1365-2966.2008.13894.x
	[arXiv:0805.0875 [astro-ph]];
	F.~Y.~Wang, Z.~G.~Dai and E.~W.~Liang,
	New Astron.\ Rev.\  {\bf 67}, 1 (2015)
	doi:10.1016/j.newar.2015.03.001
	[arXiv:1504.00735 [astro-ph.HE]].
	
	
	\bibitem{Plionis:2011jj}
	M.~Plionis, R.~Terlevich, S.~Basilakos, F.~Bresolin, E.~Terlevich, J.~Melnick and R.~Chavez,
	Mon.\ Not.\ Roy.\ Astron.\ Soc.\  {\bf 416}, 2981 (2011)
	doi:10.1111/j.1365-2966.2011.19247.x
	[arXiv:1106.4558 [astro-ph.CO]];
	;
P.~Tsiapi, S.~Basilakos, M.~Plionis, R.~Terlevich, E.~Terlevich, A.~L.~G.~Moran, R.~Chavez, F.~Bresolin, D.~F.~Arenas and E.~Telles,
[arXiv:2107.01749 [astro-ph.CO]].
	
	\bibitem{Blake:2011en}
	C.~Blake {\it et al.},
	Mon.\ Not.\ Roy.\ Astron.\ Soc.\  {\bf 418}, 1707 (2011)
	doi:10.1111/j.1365-2966.2011.19592.x
	[arXiv:1108.2635 [astro-ph.CO]];
	;
A.~de Mattia, V.~Ruhlmann-Kleider, A.~Raichoor, A.~J.~Ross, A.~Tamone, C.~Zhao, S.~Alam, S.~Avila, E.~Burtin and J.~Bautista, \textit{et al.}
Mon. Not. Roy. Astron. Soc. \textbf{501} (2021) no.4, 5616-5645
doi:10.1093/mnras/staa3891
[arXiv:2007.09008 [astro-ph.CO]].

	
	\bibitem{Ade:2015xua}
	P.~A.~R.~Ade {\it et al.} [Planck Collaboration],
	Astron.\ Astrophys.\  {\bf 594}, A13 (2016)
	doi:10.1051/0004-6361/201525830
	[arXiv:1502.01589 [astro-ph.CO]].
	\bibitem{Basilakos:2017rgc}
	S.~Basilakos and S.~Nesseris,
	Phys.\ Rev.\ D {\bf 96}, no. 6, 063517 (2017)
	doi:10.1103/PhysRevD.96.063517
	[arXiv:1705.08797 [astro-ph.CO]].
	
	\bibitem{Anagnostopoulos:2018jdq}
	F.~K.~Anagnostopoulos, S.~Basilakos, G.~Kofinas and V.~Zarikas,
	``Constraining the Asymptotically Safe Cosmology: cosmic acceleration without dark energy'',
	JCAP {\bf 1902}, 053 (2019),
	doi:10.1088/1475-7516/2019/02/053
	[arXiv:1806.10580 [astro-ph.CO]].
	
	\bibitem{Anagnostopoulos:2019mrc}
	F.~K.~Anagnostopoulos, G.~Kofinas and V.~Zarikas,
	Int. J. Mod. Phys. D \textbf{28} (2019) no.14, 14
	doi:10.1142/S0218271819440139
	[arXiv:2102.07578 [gr-qc]].
	
	

	
	
	\bibitem{c11}
	Weinberg, S. Ultraviolet divergences in quantum theories of gravitation. General Relativity: An Einstein centenary
	survey, Eds. Hawking, S.W., Israel, W; Cambridge University Press 1979, pp. 790–831.
	
	\bibitem{c12}
	Reuter, M. Nonperturbative evolution equation for quantum gravity. Phys.Rev. 1998, D57, 971–985,
	[arXiv:hep-th/hep-th/9605030]. doi:10.1103/PhysRevD.57.971.
	
	
	\bibitem{c2}
	A.~Bonanno, A.~Eichhorn, H.~Gies, J.~M.~Pawlowski, R.~Percacci, M.~Reuter, F.~Saueressig and G.~P.~Vacca,
	Front. in Phys. \textbf{8} (2020), 269
	doi:10.3389/fphy.2020.00269
	[arXiv:2004.06810 [gr-qc]].
	
	
	\bibitem{c3}
	A.~M.~Polyakov,
	[arXiv:hep-th/9304146 [hep-th]].
	
	\bibitem{c4}
	A.~Bonanno, G.~Kofinas and V.~Zarikas,
	Phys. Rev. D \textbf{103} (2021) no.10, 104025
	doi:10.1103/PhysRevD.103.104025
	[arXiv:2012.05338 [gr-qc]].
	;
	M.~Reuter and H.~Weyer,
	JCAP {\bf 0412}, 001 (2004)
	doi:10.1088/1475-7516/2004/12/001
	[hep-th/0410119];
	;
	M.~Reuter and H.~Weyer,
	Phys.\ Rev.\ D {\bf 70}, 124028 (2004)
	doi:10.1103/PhysRevD.70.124028
	[hep-th/0410117];
	;
	M.~Reuter and H.~Weyer,
	Phys.\ Rev.\ D {\bf 69}, 104022 (2004)
	doi:10.1103/PhysRevD.69.104022
	[hep-th/0311196].
	
	
	\bibitem{Reuter:2009kq}
	M.~Reuter and H.~Weyer,
	Gen.\ Rel.\ Grav.\  {\bf 41}, 983 (2009) [arXiv:0903.2971 [hep-th]];
	
	
	\bibitem{Reuter:2004nx}
	M.~Reuter and H.~Weyer,
	JCAP {\bf 0412}, 001 (2004)
	doi:10.1088/1475-7516/2004/12/001
	[hep-th/0410119];
	\bibitem{Reuter:2004nv}
	M.~Reuter and H.~Weyer,
	Phys.\ Rev.\ D {\bf 70}, 124028 (2004)
	doi:10.1103/PhysRevD.70.124028
	[hep-th/0410117];
	
	
\bibitem{Bonanno:2011yx}
A.~Bonanno and S.~Carloni,
New J. Phys. \textbf{14} (2012), 025008
doi:10.1088/1367-2630/14/2/025008
[arXiv:1112.4613 [gr-qc]].
	
	\bibitem{c5}
	A.~Bonanno and H.~E.~Fr\"ohlich,
	Astrophys. J. Lett. \textbf{893} (2020) no.2, L35
	doi:10.3847/2041-8213/ab86b9
	[arXiv:1707.01866 [astro-ph.SR]].
	
	
	\bibitem{Einstein:1946ev}
	A.~Einstein and E.~G.~Strauss,
	Annals Math.\  {\bf 47}, 731 (1946)
	doi:10.2307/1969231.
	
	
	\bibitem{Israel:1966rt}
	W.~Israel,
	Nuovo Cim.\ B {\bf 44S10}, 1 (1966)
	[Nuovo Cim.\ B {\bf 44}, 1 (1966)]
	Erratum: [Nuovo Cim.\ B {\bf 48}, 463 (1967)]
	doi:10.1007/BF02710419, 10.1007/BF02712210;
	G. Darmois, {\em M\'{e}morial des Sciences Math\'{e}matiques\/},
	Fascicule XXV (Gauthier-Villars, Paris, 1927), Chap. V.
	
	\bibitem{matching} 
	G. Darmois, {\em M\'{e}morial des Sciences Math\'{e}matiques\/},
	Fascicule XXV (Gauthier-Villars, Paris, 1927), Chap. V.;
	H. Stephani, {\em General Relativity\/} (Cambridge University
	Press, Cambridge, 1990);
	G.A. Baker, Jr. {\em Bound systems in an expanding universe\/},
	astro-ph/0003152, 10 March 2000;
	L.P. Eisenhart, {\em Riemannian Geometry\/} (Princeton
	University Press, Princeton, 1949).
	
	\bibitem{Wilson:1973jj}
	K.~G.~Wilson and J.~B.~Kogut,
	Phys. Rept. \textbf{12}, 75-199 (1974)
	doi:10.1016/0370-1573(74)90023-4.
	
	\bibitem{don}
	J.F.~Donoghue,
	Frontiers of Physics, {\bf 8},(2020) 8, [arXiv:1911.02967]; 
	J.~F.~Donoghue,
	Phys. Rev. D \textbf{50}, 3874-3888 (1994)
	doi:10.1103/PhysRevD.50.3874
	[arXiv:gr-qc/9405057 [gr-qc]].
	
	
	
	
	\bibitem{Bonanno:2000ep}
	A.~Bonanno and M.~Reuter,
	Phys. Rev. D \textbf{62}, 043008 (2000)
	doi:10.1103/PhysRevD.62.043008
	[arXiv:hep-th/0002196 [hep-th]].
	
	\bibitem{Bonanno:2019ilz}
	A.~Bonanno, R.~Casadio and A.~Platania,
	JCAP \textbf{01}, 022 (2020)
	doi:10.1088/1475-7516/2020/01/022
	[arXiv:1910.11393 [gr-qc]].
	
	\bibitem{Bonanno:2007wg}
	A.~Bonanno and M.~Reuter,
	JCAP {\bf 0708}, 024 (2007)
	doi:10.1088/1475-7516/2007/08/024
	[arXiv:0706.0174 [hep-th]].
	
	\bibitem{2016PhRvD..94j3514K}
	G.~Kofinas and V.~Zarikas,
	Phys. Rev. D \textbf{94}, no.10, 103514 (2016)
	doi:10.1103/PhysRevD.94.103514
	[arXiv:1605.02241 [gr-qc]].
	
	\bibitem{Bonanno:2017pkg}
	A.~Bonanno and F.~Saueressig,
	Comptes Rendus Physique \textbf{18}, 254-264 (2017)
	doi:10.1016/j.crhy.2017.02.002
	[arXiv:1702.04137 [hep-th]].
	
	\bibitem{Kofinas:2015sna}
	G.~Kofinas and V.~Zarikas,
	JCAP {\bf 1510}, no. 10, 069 (2015)
	doi:10.1088/1475-7516/2015/10/069;
	C.~Bambi, D.~Malafarina and L.~Modesto,
	Phys.\ Rev.\ D {\bf 88}, 044009 (2013)
	doi:10.1103/PhysRevD.88.044009
	[arXiv:1305.4790 [gr-qc]];
	D.~Malafarina,
	Universe {\bf 3}, no. 2, 48 (2017)
	doi:10.3390/universe3020048
	[arXiv:1703.04138 [gr-qc]].
	
	
\bibitem{planck}
Aghanim ~N. {\it et al.} [Planck Collaboration] 
	Astronomy \& Astrophysics {\bf 641} (2020): A6.
	doi.org/10.1051/0004-6361/201833910
	[arXiv:1807.06209 (2018)].


\bibitem{local}
R.~Jimenez, L.~Verde, T.~Treu and D.~Stern,
Astrophys. J. \textbf{593} (2003), 622-629
doi:10.1086/376595
[arXiv:astro-ph/0302560 [astro-ph]].
;
C.~Zhang, H.~Zhang, S.~Yuan, T.~J.~Zhang and Y.~C.~Sun,
Res. Astron. Astrophys. \textbf{14} (2014) no.10, 1221-1233
doi:10.1088/1674-4527/14/10/002
[arXiv:1207.4541 [astro-ph.CO]].
	
\bibitem{OHD}
Cao, S.{\it et al.} 
Universe \textbf{7.3} (2021): 57.	
doi:10.3390/universe7030057
[arXiv:2103.03670v1]

\bibitem{OHD1}
Moresco, Michele, {\it et al.}  
JCAP \textbf{2016.05} (2016): 014.
doi:10.1088/1475-7516/2016/05/014
[arXiv:1601.01701v2]

\bibitem{OHD2}
Zhang, Cong, {\it et al.}  
Res.\ Astron.\ Astrophys.\ \textbf{14.10} (2014): 1221.
doi:10.1088/1674--4527/14/10/002
[arXiv:1207.4541v3]

\bibitem{OHD3}
Jimenez, Raul, {\it et al.} 
Astrophys.\ J. \textbf{593.2} (2003): 622.
doi:10.1086/376595
[arXiv:astro-ph/0302560v1]

\bibitem{OHD4}
Simon, Joan, Licia Verde, and Raul Jimenez. 
Phys.\ Rev.\ D \textbf{71.12} (2005): 123001.
doi:10.1103/PhysRevD.71.123001
[arXiv:astro-ph/0412269v1]

\bibitem{OHD5}
Moresco, Michele, {\it et al.}  
JCAP \textbf{2012.07} (2012): 053.
doi:10.1088/1475-7516/2012/07/053
[arXiv:1201.6658v3]

\bibitem{OHD6}
Gaztanaga, Enrique, Anna Cabré, and Lam Hui. 
Mon.\ Not.\ R.\ Astron.\ Soc. \textbf{399.3} (2009): 1663-1680.
doi:10.1111/j.1365-2966.2009.15405.x
[arXiv:0807.3551v5]

\bibitem{OHD7}
Xu, Xiaoying, {\it et al.} 
Mon.\ Not.\ R.\ Astron.\ Soc. \textbf{431.3} (2013): 2834-2860.
doi:10.1093/mnras/stt379
[arXiv:1206.6732v2]

\bibitem{OHD8}
Blake, Chris, {\it et al.} 
Mon.\ Not.\ R.\ Astron.\ Soc. \textbf{425.1} (2012): 405-414.
doi:10.1111/j.1365-2966.2012.21473.x
[arXiv:1204.3674v2]

\bibitem{OHD9}
Ratsimbazafy, A. L., {\it et al.}  
Mon.\ Not.\ R.\ Astron.\ Soc. \textbf{467.3} (2017): 3239-3254.
doi:10.1093/mnras/stx301
[arXiv:1702.00418v2]

\bibitem{OHD10}
Stern, Daniel, {\it et al.}  
JCAP \textbf{2010.02} (2010): 008.
doi:10.1088/1475-7516/2010/02/008
[arXiv:0907.3149v1]


\bibitem{OHD11}
Samushia, Lado, {\it et al.}  
Mon.\ Not.\ R.\ Astron.\ Soc. \textbf{429.2} (2013): 1514-1528.
doi:	10.1093/mnras/sts443
[arXiv:1206.5309v2]

\bibitem{OHD12}
Moresco, Michele. 
Mon.\ Not.\ R.\ Astron.\ Soc. Lett. \textbf{450.1} (2015): L16-L20.
doi:10.1093/mnrasl/slv037
[arXiv:1503.01116v1]

\bibitem{OHD13}
Delubac, Timothée, {\it et al.}  
Astronomy \& Astrophysics \textbf{574} (2015): A59.
doi:10.1051/0004-6361/201423969
[arXiv:1404.1801v2]

\bibitem{OHD14}
Font-Ribera, Andreu, {\it et al.} 
JCAP \textbf{2014.05} (2014): 027.
doi:10.1088/1475-7516/2014/05/027
[arXiv:1311.1767v2]



	
\bibitem{future}
P.~S.~Apostolopoulos,
Class. Quant. Grav. \textbf{34} (2017) no.9, 095013
doi:10.1088/1361-6382/aa66df
[arXiv:1611.04569 [gr-qc]].; 
J.~D.~Barrow, S.~Basilakos and E.~N.~Saridakis,
Phys. Lett. B \textbf{815} (2021), 136134
doi:10.1016/j.physletb.2021.136134
[arXiv:2010.00986 [gr-qc]].;
G.~Aliferis and V.~Zarikas,
Phys. Rev. D \textbf{103} (2021) no.2, 023509
doi:10.1103/PhysRevD.103.023509
[arXiv:2006.13621 [gr-qc]].;
Y.~Eroshenko,
Phys. Dark Univ. \textbf{32} (2021), 100833
doi:10.1016/j.dark.2021.100833
[arXiv:2105.03704 [astro-ph.CO]].

\bibitem{sn1}
 Linder, Eric V. and Ayan Mitra  Phys.\ Rev.\ D\ {\bf100} (2019): 043542.
doi:10.1103/PhysRevD.100.043542
[arXiv:1907.00985v1]


\bibitem{sn2}
Mitra, Ayan and Eric V. Linder.  Phys.\ Rev.\ D\ {\bf103.2} (2021): 023524.
doi:10.1103/PhysRevD.103.023524
[arXiv:2011.08206v2]

\bibitem{FRB}
Lau, Albert Wai Kit {\it et al.}.
New\ Astronomy\ {\bf 89} (2021): 101627.
doi:https:10.1016/j.newast.2021.101627
[arXiv:2006.11072v2]

\bibitem{gw}
Mitra, Ayan,{\it et al.}.  Mon.\ Not.\ R.\ Astron.\ Soc. Lett. {\bf 502.4} (2021): 5563-5575.
doi:10.1093/mnras/stab165
[arXiv:2010.00189v1]

\end{thebibliography}
\end{document}